\documentclass[letterpaper, 10 pt, conference]{ieeeconf}  % Comment this line out if you need a4paper

\IEEEoverridecommandlockouts                              % This command is only needed if 
\usepackage{xcolor}
 
\usepackage{graphicx,times}
\usepackage{amsmath,amssymb}
\usepackage{listings}
\usepackage{multirow}
\usepackage{hyperref}

\title{\LARGE \bf
Classification of Influenza Hemagglutinin Protein Sequences using Convolutional Neural Networks
}

\author{Charalambos Chrysostomou$^{1*}$, Floris Alexandrou$^{1}$, Mihalis A. Nicolaou$^{1}$ and Huseyin Seker$^{2}$% <-this % stops a space
\thanks{$^{1}$Charalambos Chrysostomou, Floris Alexandrou and Mihalis Nicolaou are with Computation-based Science and Technology Research Center, The Cyprus Institute, Nicosia, Cyprus
        {\tt\small c.chrysostomou@cyi.ac.cy, f.alexandrou@cyi.ac.cy, m.nicolaou@cyi.ac.cy}}%
\thanks{$^{2}$Huseyin Seker is with the Faculty of Computing, Engineering and the Built Environment, Birmingham City University, Birmingham, B5 5JU The United Kingdom
        {\tt\small hseker1970@yahoo.co.nz}}%
\thanks{$^{*}$ Corresponding Author}%
}

\begin{document}

\maketitle
\thispagestyle{empty}
\pagestyle{empty}

%%%%%%%%%%%%%%%%%%%%%%%%%%%%%%%%%%%%%%%%%%%%%%%%%%%%%%%%%%%%%%%%%%%%%%%%%%%%%%%%
\begin{abstract}
The Influenza virus can be considered as one of the most severe viruses that can infect multiple species with often fatal consequences to the hosts. The Hemagglutinin (HA) gene of the virus can be a target for antiviral drug development realised through accurate identification of its sub-types and possible the targeted hosts.  This paper focuses on accurately predicting if an Influenza type A virus can infect specific hosts, and more specifically, Human, Avian and Swine hosts, using only the protein sequence of the HA gene.  In more detail, we propose encoding the protein sequences into numerical signals using the Hydrophobicity Index and subsequently utilising a Convolutional Neural Network-based predictive model.  The  Influenza HA protein sequences used in the proposed work are obtained from the Influenza Research Database (IRD).  Specifically, complete and unique HA protein sequences were used for avian, human and swine hosts. The data obtained for this work was 17999 human-host proteins, 17667 avian-host proteins and 9278 swine-host proteins. Given this set of collected proteins, the proposed method yields as much as 10\% higher accuracy for an individual class (namely, Avian) and 5\% higher overall accuracy than in an earlier study. It is also observed that the accuracy for each class in this work is more balanced than what was presented in this earlier study. As the results show, the proposed model can distinguish HA protein sequences with high accuracy whenever the virus under investigation can infect Human, Avian or Swine hosts.

\end{abstract}

%%%%%%%%%%%%%%%%%%%%%%%%%%%%%%%%%%%%%%%%%%%%%%%%%%%%%%%%%%%%%%%%%%%%%%%%%%%%%%%%
\section{INTRODUCTION}

Pathogens, including viruses, can cause infectious diseases to spread within populations. These pathogens can be transmitted in multiple ways, with high transmission rates in most circumstances. Identification and diagnosis of infectious diseases are vital, especially in novel pathogens, to prevent and control global pandemics, such as the current COVID-19 pandemic \cite{nicola2020socio}.

Influenza viruses are part of the Orthomyxoviridae family of viruses that have negative-sense, single-stranded, segmented RNA genomes, with the majority of the virus burden being associated with influenza viruses type A and B \cite{zambon1999epidemiology}. Influenza viruses capable of infecting humans were introduced from birds and swine \cite{russell2014science}. Their introduction to humans has begun global pandemics with the 1918 "Spanish flu" and 2009 "Swine flu" pandemics. Influenza viruses are responsible for more than 500,000 deaths worldwide and affect around 5–15\% of the population each year \cite{stohr2002influenza}. The evolution of influenza viruses enables them to infect individuals who have previously gained immunity through vaccination or previous infections. Furthermore, Influenza viruses can be effectively transmitted within populations from direct contact, respiratory droplets, objects, or materials in the environment, such as clothes, utensils, and furniture. Currently, vaccines are the most efficient and practical method in limiting the influenza virus's spread and stop Influenza epidemics. These vaccines must be renewed periodically to maintain their effectiveness against Influenza viruses. 

The Influenza virus genome contains eight gene segments that include: PB2 (polymerase subunit) responsible for encoding RNA; PB1 (polymerase subunit) responsible for encoding RNA and includes the PB1-F2 protein, which causes cell death; PA (polymerase subunit) responsible for encoding RNA and includes the PA-X protein, responsible for host transcription shutoff function; NP (nucleoprotein); M1 and M2 (matrix proteins); NS1 and NEP (non-structural proteins); NA (neuraminidase) which facilitates the release of new viruses from the infected host cell; HA (hemagglutinin) responsible for binding and entry of the virus to the host cell. As this study's primary aim is to classify the capability of an Influenza virus to infect different hosts, we will focus primarily on the HA gene. Influenza viruses are classified into subtypes based on the organisation of HA and NA glycoproteins on their surfaces. Currently, 18 HA subtypes and 11 NA subtypes exist in the wild \cite{tong2013new} with the majority affecting Avian species.  Only three subtypes are known to adapt and circulate in Humans H1N1, H2N2 and H3N2. Seasonal epidemics are caused by two of these Influenza subtypes H1N1 and H3N2.

In the literature, various methods were developed and used for analysing and characterising protein sequences. More specifically, for the classification and characterisation of Influenza subtypes, \cite{chrysostomou2017prediction, chrysostomou2010complex, carmona2013fuzzy, chrysostomou2011effects, chrysostomou2015cisaps, chrysostomou2016structural}, methods such as Complex resonant recognition model in analysing influenza subtype protein sequences \cite{chrysostomou2010complex}, CISAPS: Complex informational spectrum for the analysis of protein sequences \cite{chrysostomou2015cisaps}, and Structural Classification of protein sequences based on signal processing and support vector machines \cite{chrysostomou2016structural}, have been proposed and developed. 

In this paper, a new method is proposed based on deep learning methodologies and, more specifically, Convolutional Neural Networks (CNN) to classify the Influenza protein sequences based on their capability to infect the specific host by utilising only the amino acid sequence of the protein sequence to accomplish unprecedented accuracy. The paper is organised as follows: Section \ref{section:methodsandmaterials} presents the methods and materials developed and used, while Section \ref{section:results} presents the results obtained. Finally, concluding remarks are outlined in Section \ref{section:conclusions}.

\section{Methods and Materials}
\label{section:methodsandmaterials}

\subsection{Protein Sequences}

\begin{table}
\normalsize
\renewcommand{\arraystretch}{1.2}
\caption{Number of HA Protein Sequences}
\centering
\begin{tabular}{c c c c}
\hline
 HA Subtype & Human & Avian & Swine\\
 \hline
H1* & 8426 & 594 & 6626\\
H2* & 81 & 416 & 2\\
H3* & 9134 & 1316 & 2429\\
H4* & 0 & 1001 & 5\\
H5* & 220 & 3653 & 24\\
H6* & 0 & 1485 & 2\\
H7* & 100 & 1310 & 3\\
H8* & 0 & 149 & 0\\
H9* & 14 & 5009 & 20\\
H10* & 2 & 621 & 1\\
H11* & 0 & 654 & 1\\
H12* & 0 & 290 & 0\\
H13* & 0 & 276 & 0\\
H14* & 0 & 19 & 0\\
H15* & 0 & 10 & 0\\
H16* & 0 & 140 & 0\\
mixed & 22 & 724 & 165\\
\hline
\textbf{Total} & \textbf{17999} & \textbf{17667} & \textbf{9278}\\
\hline
\end{tabular}
\label{number_of_HA_proteins}
\end{table}

\begin{figure*}[ht]
    \centering
    \includegraphics[width=1.0\textwidth, height=0.35\textheight]{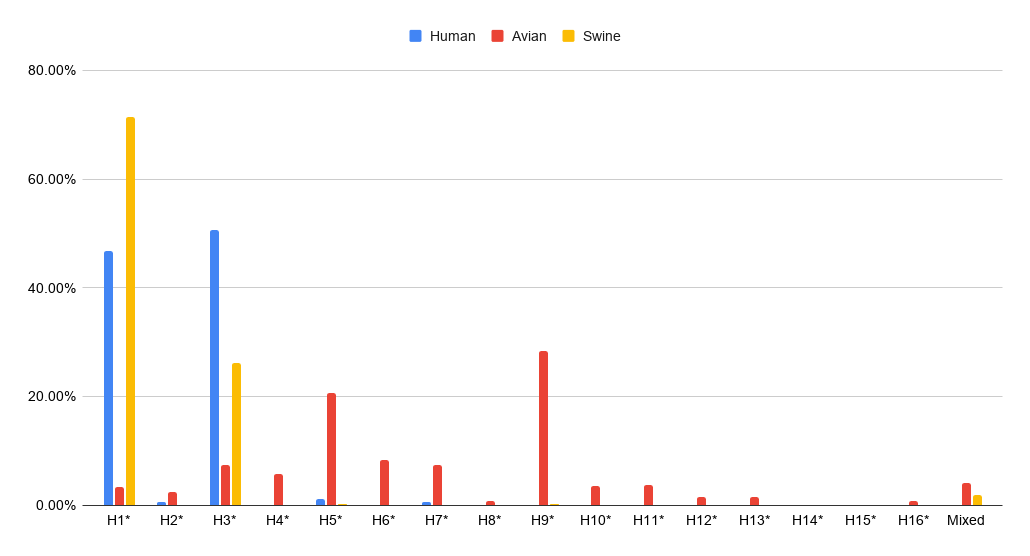}
    \caption{Percentage of HA proteins per class and species}
    \label{HA_protein_sequences_figure}
\end{figure*}

\begin{table}[ht]
\renewcommand{\arraystretch}{1.2}
\normalsize
\caption{Hydrophobicity Amino Acid Index used to encode protein sequences from alphabetical to numerical characters for analysis}
\setlength{\tabcolsep}{6pt}
\centering 
\begin{tabular}{l c c} 
\hline 

\multirow{2}{*}{Amino acid} & \multicolumn{2}{l}{Hydrophobicity Values} \\
                            & Original           & Normalised          \\
\hline
Alanine (A) &0.61&0.23\\
Arginine (R) &0.60&0.23\\
Asparagine (N) &0.06&0.02\\
Aspartic acid (D) &0.46&0.17\\
Cysteine (C) &1.07&0.40\\
Glutamine (Q) &0.00&0.00\\
Glutamic acid (E) &0.47&0.18\\
Glycine (G) &0.07&0.03\\
Histidine (H) &0.61&0.23\\
Isoleucine (I) &2.22&0.84\\
Leucine (L) &1.53&0.58\\
Lysine (K) &1.15&0.43\\
Methionine (M) &1.18&0.45\\
Phenylalanine (F) &2.02&0.76\\
Proline (P) &1.95&0.74\\
Serine (S) &0.05&0.02\\
Threonine (T) &0.05&0.02\\
Tryptophan (W) &2.65&1.00\\
Tyrosine (Y) &1.88&0.71\\
Valine (V)  &1.32&0.50\\
Other&0.00&0.00\\

  \hline 
\end{tabular} 
\label{hydro_values} 
\end{table}

The Influenza HA protein sequences used in the proposed work are obtained from The Influenza Research Database (IRD) \cite{zhang2017influenza}. Specifically, complete and unique Hemagglutinin (HA) protein sequences were used for avian, human and swine hosts, which are the primary hosts affected with the virus. The data obtained for this work was 17999 human-host proteins, 17667 avian-host proteins and 9278 swine-host proteins. The complete list of proteins, including HA subclasses, used in this study can be found in Table \ref{number_of_HA_proteins}, and Figure \ref{HA_protein_sequences_figure} illustrates the percentage of HA proteins per class and species.

The proposed analysis was performed directly to the protein sequence using a Hydrophobicity index \cite{argos1982structural} to encode the protein sequences from alphabetical to numerical characters for analysis. Before encoding the protein sequences, the amino acid index values were re-normalised to 0-1. Any character beyond the standard 20 amino acids was encoded using the value 0. The complete list of the Hydrophobicity amino acid index can be found in Table \ref{hydro_values}.

The collected protein sequences have variable sizes, with 576 being the maximum number of amino acids. To encode the sequences and use the proposed model, proteins with a lower number of amino acids were padded with 0's at the end of the sequence to reach the mentioned number. The data was further encoded with one-hot encoding for the multi-class classification problem.

\subsection{Proposed Model}
\begin{figure*}[ht]
    \centering
    \includegraphics[width=0.99\textwidth, height=0.18\textheight]{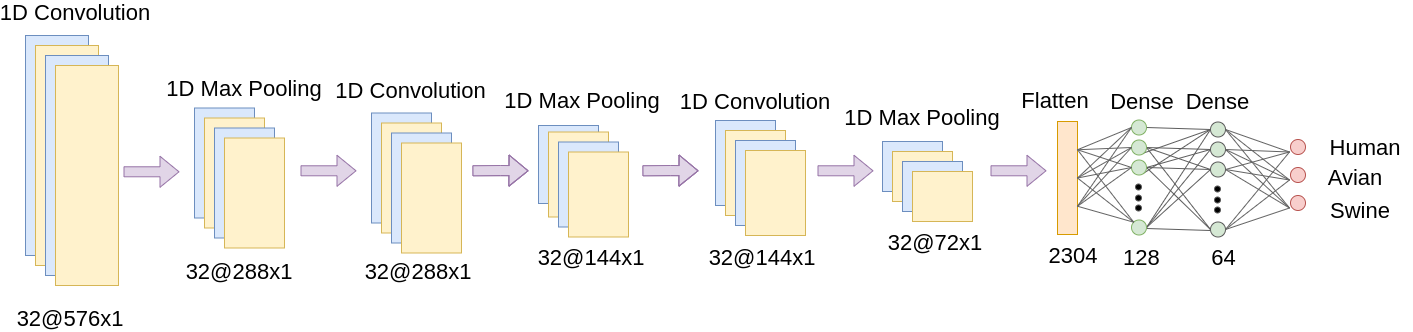}
    \caption{Structure of the Proposed Model based on Convolutional Neural Networks}
    \label{NN_figure}
\end{figure*}

The proposed work is based on a Convolutional Neural Network (CNN). CNN's are a subtype of deep feed-forward artificial neural networks that have been applied and used to analyse visual representations but recently have been used in other domains, including classification and characterisation of protein sequences \cite{zeng2016convolutional, wang2016protein, zhao2020identification}. CNN's utilise a variation of multi-layer perceptrons created to minimise the pre-processing required \cite{goodfellow2016deep}, thus in comparison, CNN's require relatively minimal pre-processing of data in relation to other classification methodologies and translates to a substantial advantage where prior knowledge and expertise of any given problem is not available or unknown.

As shown in Figure \ref{NN_figure}, the proposed model structure consists of three convolutional layers of 32, 3x3 kernels, followed by a max-pooling layer and a rectified nonlinear activation function (Leaky ReLU) \cite{maas2013rectifier} transforming the feature space from 32x576 to 32x72. The fourth block consists of a Flatten layer that transforms the feature space from 32x72 to 2304x1. The fifth and sixth layers are fully connected layers of 128 and 64 neurons, respectively, with Leaky ReLU as the activation function. The output layer consists of 3 neurons that correspond to each of the available species and utilises the softmax activation function \cite{goodfellow2016deep}. The model is trained using the “Adam” optimiser \cite{kingma2014adam} and the “categorical cross-entropy” loss function \cite{goodfellow2016deep}. The proposed model and hyperparameters were selected based on a trial and error process to maximise the training and validation accuracy. The proposed methodology source code is available at \url{https://gitlab.com/charalambos.chrysostomou/embc21_influenza.git}

\section{Results and Discussion}
\label{section:results}

% \addtolength{\textheight}{-0.5cm}

\begin{table*}
\setlength{\tabcolsep}{15pt}
\renewcommand{\arraystretch}{1.3}

\normalsize
\caption{Results}
\centering

\begin{tabular}{lccc}
Set        & Accuracy  & MCC Score & F1 Score \\
\hline
Training   & 99.36\% $\pm$ 0.11\% & 0.990 $\pm$ 0.002 & 0.994 $\pm$ 0.001 \\
Validation & 98.84\% $\pm$ 0.20\% & 0.982 $\pm$ 0.003 & 0.988 $\pm$ 0.002 \\
Test       & 98.78\% $\pm$ 0.22\% & 0.981 $\pm$ 0.003 & 0.988 $\pm$ 0.002 \\
\hline

\end{tabular}
\label{table_results1}
\end{table*}

\begin{table}
\setlength{\tabcolsep}{15pt}
\renewcommand{\arraystretch}{1.3}

\normalsize
\caption{Results based on the Test sets per Species}
\centering

\begin{tabular}{lc}
Set        & Accuracy \\
\hline
Human   & 98.74\% $\pm$ 0.32\% \\
Avian & 99.52\% $\pm$ 0.25\% \\
Swine      & 97.19\% $\pm$ 0.75\% \\
\hline

\end{tabular}
\label{table_results2}
\end{table}

In this paper, a classification model is presented, based on Deep Learning and Convolutional Neural Networks (CNN), for the characterisations and classification Influenza type A based upon the ability to infect a specific host, more specifically human, avian and swine hosts, by solely using the HA protein sequence. 

To ensure that the proposed model is accurate and the results can be generalised, the model was trained 100 times for 1000 epochs, with different random subsets chosen uniformly and assigned to training, validation and testing sets. For each training cycle, the best weights were saved based on the validation error. To evaluate the proposed method's efficacy, the average of the test set, for the total accuracy, Matthews correlation coefficient (MCC) \cite{matthews1975comparison} and F-score (F1) \cite{sasaki2007truth} scores, were considered.  The predictive model's average accuracy is 99.36\% $\pm$ 0.11\%, 98.84\% $\pm$ 0.20\% and 98.78\% $\pm$ 0.22\% for the training, validation and test sets, respectively. For the MCC score, the results are 0.990 $\pm$ 0.002,  0.982 $\pm$ 0.003 and 0.981 $\pm$ 0.003 for the training, validation and test sets, respectively. Finally, for the F1 score, the results are 0.994 $\pm$ 0.001, 0.988 $\pm$ 0.002 and 0.988 $\pm$ 0.002 for the training, validation and test sets, respectively. Detailed results can be found in Tables \ref{table_results1} and \ref{table_results2}. The results based on the test sets per species were also calculated with 98.74\% $\pm$ 0.32\%, 99.52\% $\pm$ 0.25\% and 97.19\% $\pm$ 0.75\% for Human, Avian and Swine species, respectively. The proposed method yields almost 10\%, 5\% and 2\% higher accuracy for Avian, Human and Swine, respectively, than those of the earlier study \cite{sherif2017classification}. It is also observed that the accuracy for each class presented in our work is more balanced than what was presented in this earlier study. The final overall accuracy is also found to be as much as 5\% higher than that of the earlier study.

\section{CONCLUSIONS}
\label{section:conclusions}

The paper presents a highly successful predictive model to classify Influenza viruses based on their capability to infect different hosts, including Human, Avian and Swine Hosts, as the results show compared to existing methods. For this study, the protein sequence of the HA gene was used, which is responsible for attaching to the host's cell and can be considered a promising antiviral drug candidate. The collected protein sequences were encoded using a normalised hydrophobicity amino acid index.

As published in the literature, more than 600 unique amino acid indices exist that describes a physicochemical feature of the protein \cite{chrysostomou2015cisaps}. Future studies are needed to identify potential alternative amino acid index or set of indices capable of better representing HA proteins and delivering even more reliable results. As the recent events of the COVID-19 pandemic have shown, a computational tool capable of identifying novel and potentially dangerous viruses in the wild that have acquired the capability to infect Human hosts will be crucial and needed to predict and control future outbreaks.

% \addtolength{\textheight}{-12cm}   % This command serves to balance the column lengths
                                  % on the last page of the document manually. It shortens
                                  % the textheight of the last page by a suitable amount.
                                  % This command does not take effect until the next page
                                  % so it should come on the page before the last. Make
                                  % sure that you do not shorten the textheight too much.

%%%%%%%%%%%%%%%%%%%%%%%%%%%%%%%%%%%%%%%%%%%%%%%%%%%%%%%%%%%%%%%%%%%%%%%%%%%%%%%%

%%%%%%%%%%%%%%%%%%%%%%%%%%%%%%%%%%%%%%%%%%%%%%%%%%%%%%%%%%%%%%%%%%%%%%%%%%%%%%%%

%%%%%%%%%%%%%%%%%%%%%%%%%%%%%%%%%%%%%%%%%%%%%%%%%%%%%%%%%%%%%%%%%%%%%%%%%%%%%%%%
% \section*{APPENDIX}

% Appendixes should appear before the acknowledgment.

% \section*{ACKNOWLEDGMENT}

% The preferred spelling of the word ÒacknowledgmentÓ in America is without an ÒeÓ after the ÒgÓ. Avoid the stilted expression, ÒOne of us (R. B. G.) thanks . . .Ó  Instead, try ÒR. B. G. thanksÓ. Put sponsor acknowledgments in the unnumbered footnote on the first page.

%%%%%%%%%%%%%%%%%%%%%%%%%%%%%%%%%%%%%%%%%%%%%%%%%%%%%%%%%%%%%%%%%%%%%%%%%%%%%%%%

\bibliographystyle{IEEEtran}
\bibliography{ref.bib}

\end{document}